# Low damping and microstructural perfection of sub-40nm-thin yttrium iron garnet films grown by liquid phase epitaxy

Carsten Dubs,[1*] Oleksii Surzhenko,[1] Ronny Thomas,[2] Julia Osten,[2] Tobias Schneider,[2] Kilian Lenz,[2] Jörg Grenzer,[2] René Hübner,[2] Elke Wendler[3]

[1] INNOVENT e.V. Technologieentwicklung, Prüssingstr. 27B, 07745 Jena, Germany
[2] Institute of Ion Beam Physics and Materials Research, Helmholtz-Zentrum Dresden-Rossendorf, Bautzner Landstr. 400, 01328 Dresden, Germany
[3] Institut für Festkörperphysik, Friedrich-Schiller-Universität Jena, Helmholtzweg 3, 07743 Jena, Germany
[*] Correspondence: cd@innovent-jena.de

The field of magnon spintronics is experiencing an increasing interest in the development of solutions for spin-wave-based data transport and processing technologies that are complementary or alternative to modern CMOS architectures. Nanometer-thin yttrium iron garnet (YIG) films have been the gold standard for insulator-based spintronics to date, but a potential process technology that can deliver perfect, homogeneous large-diameter films is still lacking. We report that liquid phase epitaxy (LPE) enables the deposition of nanometer-thin YIG films with low ferromagnetic resonance losses and consistently high magnetic quality down to a thickness of 20 nm. The obtained epitaxial films are characterized by an ideal stoichiometry and perfect film lattices, which show neither significant compositional strain nor geometric mosaicity, but sharp interfaces. Their magneto-static and dynamic behavior is similar to that of single crystalline bulk YIG. We found, that the Gilbert damping coefficient $\alpha$ is independent of the film thickness and close to $1 \times 10^{-4}$, and that together with an inhomogeneous peak-to-peak linewidth broadening of $\Delta H_{0\parallel} = 0.4$ G, these values are among the lowest ever reported for YIG films with a thickness smaller than 40 nm. These results suggest, that nanometer-thin LPE films can be used to fabricate nano- and micro-scaled circuits with the required quality for magnonic devices. The LPE technique is easily scalable to YIG sample diameters of several inches.

## I. INTRODUCTION

Yttrium iron garnet ($Y_3Fe_5O_{12}$; YIG) in the micrometer thickness range is the material of choice in radio-frequency (RF) engineering for decades (see, e.g., Refs. [1-5]). Especially the lowest spin wave loss of all known magnetic materials and the fact, that it is a dielectric are of decisive importance. Since one has learned how to grow YIG films in the nanometer thickness range, there has been a renaissance of this material, as its magnetic and microwave properties are in particular demand in many areas of modern physics.

A growing field of application for magnetic garnets is (i) magnonics, which deals with future potential devices for data transfer and processing using spin waves [1,6-9]. The significant thickness reduction achieved today allows reducing the circuit sizes from classical millimeter dimensions [1] down to 50 nm [10-12]. Another important field is (ii) spintronics: By increasing the YIG surface-to-volume ratio as much as possible (while keeping its magnetic properties), physical phenomena, such as the inverse spin Hall effect [13], spin-transfer torque [14], and the spin Seebeck effect [15] (generated by a spin angular momentum transfer at the interfaces between YIG and a nonmagnetic metallic conductor layer) become much more efficient [7,16-29]. Also (iii) the field of terahertz physics, which uses ultrafast spin dynamics to control ultrafast magnetism, for example for potential terahertz spintronic devices [30,31,32], and (iv) the field of low-temperature physics, which deals with magnetization dynamics at cryogenic temperatures [33] for prospective quantum computer systems, are possible fields of applications for nanometer-thin iron garnet films.

There are several different techniques to grow YIG on different substrates. (i) Pulsed laser deposition (PLD) is an excellent technique for fabricating small samples of nanometer-thin YIG films with narrow ferromagnetic resonance (FMR) linewidths [17,19,21,22,28,34-36] whereas its up-scaling to larger sample dimensions of several inches is challenging. (ii) Magnetron sputtered YIG usually yields wider FMR linewidths, and inhomogeneous line broadening is frequently observed [37-40]. (iii) For large-scale, low-cost chemical solution techniques, such as spin coating, strongly broadened FMR linewidths and increased Gilbert damping parameters were reported [41,42]. (iv) Liquid phase epitaxy (LPE) from high-temperature solutions (flux melts), is a well-established technique. Since nucleation and crystal growth take place under almost thermodynamic equilibrium conditions, this guarantees high quality with respect to narrow absolute FMR linewidths and a small Gilbert damping coefficient [43-45] at the same time, making LPE comparable or superior to the other growth techniques. In addition, LPE allows YIG to be deposited in the required quality on 3- or 4-inch wafers [46]. This is important for possible applications mentioned above.

So far, classical LPE was applied to grow micrometer-thick samples used for magneto-static microwave devices [47,48] or for magneto-optical imaging systems [49]. The typical shortcomings of the LPE technology making thin-film growth so difficult lie in the fact, that, due to high growth rates, nanometer-thin films were technologically difficult to access. The etch-back processes in high-temperature solutions or interdiffusion processes at the substrate/film interface at high temperatures usually prevent sharp interfaces. In addition, film contamination by flux melt constituents (if it is not a self-flux without foreign components) is unavoidable in most cases. Nevertheless, it was recently demonstrated, that epitaxial films of 100 nm or thinner are also accessible with this technique [50,51].

In this study, we will show that we are able to deposit nanometer-thin YIG LPE films with low FMR losses and consistently high magnetic quality down to a thickness of 20 nm. There is no thinnest "ultimate" thickness for iron garnet LPE films, as it is sometimes claimed.

It should be pointed out, that, in addition to the damping properties, magnetic anisotropy contributions as a function of the sample stoichiometry and film/substrate pairing are also of great importance, since they determine the static and dynamic magnetization of the epitaxial iron garnet films and thus their possible applications. For example, large negative uniaxial anisotropy fields were usually observed for garnet films under compression, such as for YIG on gadolinium gallium garnet ($Gd_3Ga_5O_{12}$; GGG) or other suitable substrates with smaller lattice parameters grown by gas phase deposition techniques (see e.g. Refs. [35,36,52-57]), which favors in-plane magnetization. Large perpendicular magnetic anisotropies, on the other hand, can be found for films under tensile strain, e.g. on substrates with larger lattice parameter or for rare earth iron garnet films with smaller lattice parameter than GGG (see e.g. Refs. [58-62]). Between these two extremes are YIG LPE films, which are usually grown on standard GGG substrates and exhibit small tensile strain if no lattice misfit compensation, e.g. by La ion substitution [50,63], has been performed. Such films are characterized by a small uniaxial magnetic anisotropy and dominant shape anisotropy when no larger growth–induced anisotropy contributions due to Pb or Bi substitution occurs [64].

However, only little information about the structural properties and the thickness-dependent magnetic anisotropy contributions of nanometer-thin LPE films has been published so far, which is why we are concentrating on these properties for YIG films with thicknesses down to 10 nm. This allowed us to describe the intrinsic damping behavior over a wide frequency range and to determine a set of magnetic anisotropy parameters for all investigated films.

## II. EXPERIMENTAL DETAILS

Nanometer-thin YIG films were deposited on 1-inch (111) GGG substrates by LPE from $PbO-B_2O_3$-based high-temperature solutions (HTL) at about 865°C using the isothermal dipping method (see e.g. [65]). Nominally pure $Y_3Fe_5O_{12}$ films with smooth surfaces were obtained within one minute deposition time on horizontally rotated substrates with rotation rates of 100 rpm. The only variable growth parameter for all samples in this study was the degree of undercooling ($\Delta T = T_L - T_{epitaxy}$) that was restricted to $\Delta T \leq 5$ K to obtain films with thicknesses between 10 and 110 nm. Here $T_L$ is the liquidus temperature of the high-temperature solution and $T_{epitaxy}$ is the deposition temperature. After deposition, the samples were pulled out of the solution followed by a spin-off of most of the liquid melt remnants at 1000 rpm, pulled out of the furnace and cooled down to room temperature. Subsequently, the sample holder had to be stored with the sample in a diluted, hot nitric-acetic-acid solution to remove the rest of the solidified solution residues. Finally, the reverse side YIG film of the doubled-sided grown samples was removed by mechanical polishing and samples were cut into chips of different sizes by a diamond wire saw. The film thicknesses were determined by X-ray reflectometry (XRR) and by high-resolution X-ray diffraction (HR-XRD) analysis, and the latter data were used to calculate anisotropy and magnetization values.

Atomic force microscopy (AFM) using a Park Scientific M5 instrument was carried out for each sample at three different regions over 400 $\mu m^2$ ranges to determine the root-mean-square (RMS) surface roughness.

The XRR measurements were carried out using a PANanalytical/X-Pert Pro system. For the HR-XRD investigations, a Seifert-GE XRD3003HR diffractometer using a point focus was equipped with a spherical 2D Göbel mirror and a Bartels monochromator on the source side. Both systems use $Cu_{K\alpha 1}$ radiation. Reciprocal space maps (RSMs) were measured with the help of a position-sensitive detector (Mythen 1k) at the symmetric (444) and (888) as well as the asymmetric (088), (624), and (880) reflections. To obtain the highest possible angular resolution for symmetric $\theta-2\theta$ line scans, a triple-axis analyzer in front of a scintillation counter was installed on the detector. Using a recursive dynamical algorithm implemented in the commercial program RC_REF_Sim_Win [66], the vertical lattice misfits were calculated.

Rutherford backscattering spectrometry (RBS) was applied to investigate the composition of the grown YIG films using 1.8 MeV He ions and a backscattering angle of 168°. Backscattering events were registered with a common Si detector. The energy calibration of the multichannel analyzer revealed 3.61 keV per channel. A thin carbon layer was deposited on top of the samples to avoid charging during analysis. The samples were tilted by 5° with respect to the incoming He ion beam and rotated around the axis perpendicular to the sample surface in order to obtain reliable random spectra. The analysis of the measured spectra was performed by a home-made software [67] based on the computer code NDF [68] and then enables the calculation of the RBS spectra. The measured data were fitted by calculated spectra to extract the film composition. In this way, the Fe-to-Y ratio of the films was determined. Because of the low mass of oxygen, the O signal of the deposited films is too low for quantitative analysis.

High-resolution transmission electron microscopy (HR-TEM) investigations were performed with an image $C_s$-corrected Titan 80-300 microscope (FEI) operated at an accelerating voltage of 300 kV. High-angle annular dark-field scanning transmission electron microscopy (HAADF-STEM) imaging and spectrum imaging analysis based on energy-dispersive X-ray spectroscopy (EDXS) were done at 200 kV with a Talos F200X microscope equipped with a Super-X EDXS detector system (FEI). Prior to TEM analysis, the specimen mounted in a high-visibility low-background holder was placed for 10 s into a Model 1020 Plasma Cleaner (Fischione) to remove possible contaminations. Classical cross-sectional TEM-lamella preparation was done by sawing, grinding, polishing, dimpling, and final Ar-ion milling. Quantification of the element maps including Bremsstrahlung background correction based on the physical TEM model, series fit peak deconvolution, and application of tabulated theoretical Cliff-Lorimer factors as well as absorption correction was done for the elements Y ($K_\alpha$ line), Fe ($K_\alpha$ line), Gd ($L_\alpha$ line), Ga ($K_\alpha$ line), O (K line), and C (K line) using the ESPRIT software version 1.9 (Bruker).

The ferromagnetic resonance (FMR) absorption spectra were taken on two different setups. The frequency-swept measurements were recorded on a Rohde & Schwarz ZVA 67 vector network analyzer attached to a broadband stripline. The YIG/GGG sample was mounted face-down on the stripline, and the transmission signals $S_{21}$ and $S_{12}$ were recorded using a source power of -10 dBm (= 0.1 mW). The microwave frequency was swept across the resonance frequency $f_{res}$, while the in-plane magnetic field $H$ remained constant. Each recorded frequency spectrum was fitted by a Lorentz function and allowed us to define the resonance frequency $f_{res}$ and the frequency linewidth $\Delta f_{FWHM}$ corresponding to the applied field $H = H_{res}$.

In addition, field-swept measurements were carried out with another setup using an Agilent E8364B vector network analyzer and an 80-µm-wide coplanar waveguide. Again, the microwave transmission parameter $S_{21}$ was recorded as the FMR signal. This time, the microwave frequency was kept constant and the external magnetic field was swept through resonance. This facilitates tracking the FMR signals over large frequency ranges. The microwave power was set to 0 dBm (= 1 mW). In addition, this setup allowed for azimuthal and polar angle-dependent measurements to determine the anisotropy and damping contributions in detail. The FMR spectra were fitted by a complex Lorentz function to retrieve the resonance field $H_{res}$ and field-swept peak-to-peak linewidth $\Delta H_{pp}$. By fitting the four sets of resonance field data, i.e. (i) the in-plane and (ii) the perpendicular-to-plane frequency dependence as well as (iii) the azimuthal and (iv) polar angular dependences at $f$ = 10 GHz, with the resonance equation for the cubic (111) garnet system, a consistent set of anisotropy parameters was determined for each sample. In addition, the damping parameters and contributions were determined from the frequency- and angle-dependent linewidth data.

The vibrating sample magnetometer (VSM, MicroSense LLC, EZ-9) was used to measure the magnetic moments of the YIG/GGG samples magnetized along the YIG film surface. The external magnetic field $H$ was controlled within an error of ≤0.01 Oe. To estimate the volume magnetization $M$ of the YIG films, the raw VSM signal was corrected from background contributions (due to the sample holder and the GGG substrate) and normalized to the YIG volume. The Curie temperatures $T_C$ for the YIG samples were determined by zero-extrapolation of the temperature dependencies $M$ ($H$=const, $T$) measured in small in-plane magnetic fields. In order to verify the Curie temperatures measured by VSM, a differential thermal analysis of a 0.55 mm thick YIG single crystal slice was carried out and then used as a reference sample for the VSM temperature calibration.

## III. RESULTS AND DISCUSSION

### A. Microstructural properties of nanometer-thin YIG films

The thickness values reported in this study are derived from the Laue oscillations observed in the $\theta$-$2\theta$ patterns of the high-resolution X-ray diffraction (HR-XRD) measurements and are confirmed by X-ray reflectivity (XRR) measurements (see Fig. 1). The differences between both methods for determining the film thickness are in the range of ±1 nm. The surface roughness of the films, measured by atomic force microscopy (AFM) reveals RMS values ranging between 0.2 and 0.4 nm, independent of the film thickness. Sometimes, however, partial remnants of dendritic overgrowth increase the surface roughness to RMS values above 0.4 nm for inspection areas larger than 400 μm$^2$ (see, e.g., the disturbance in the top right corner of the AFM image inset in Fig. 1).

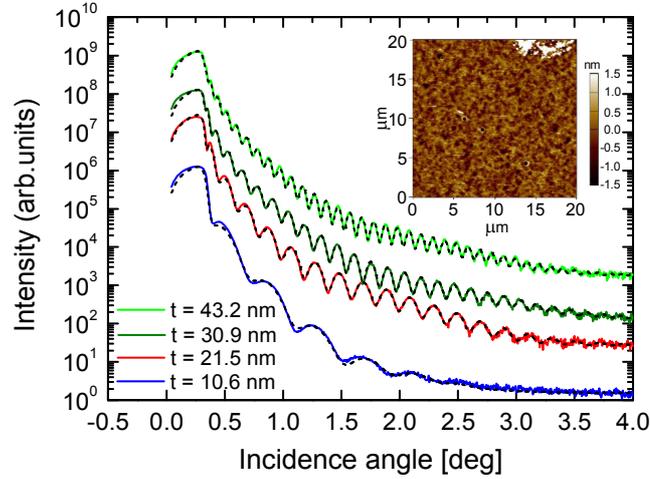

FIG. 1. XRR plots of LPE-grown YIG films of different thicknesses. Solid lines correspond to the experimental data, while dashed lines represent the fitted curves. The spectra shifted vertically for ease of comparison. The inset shows an AFM image of the surface topography of the 11 nm YIG film with a RMS roughness of 0.4 nm.

*1. Epitaxial perfection studied by high-resolution X-ray diffraction*

Combined high-resolution reciprocal space map (HR-RSM) investigations around asymmetric and symmetric Bragg reflections are useful to evaluate the intergrowth relations of epitaxial films on single-crystalline substrates as well as to distinguish between lattice strain induced by the film lattice distortion or compositional changes due to stoichiometric deviations.

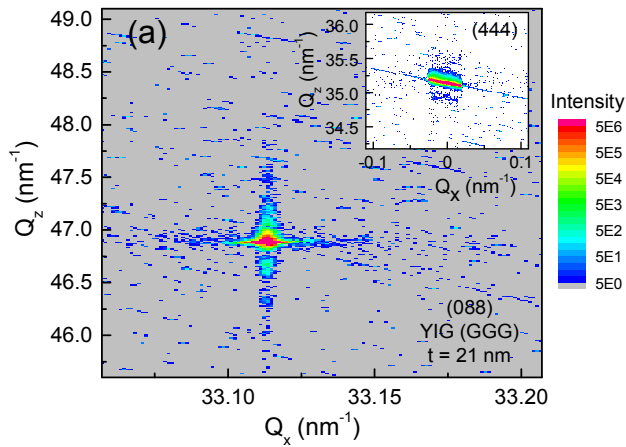

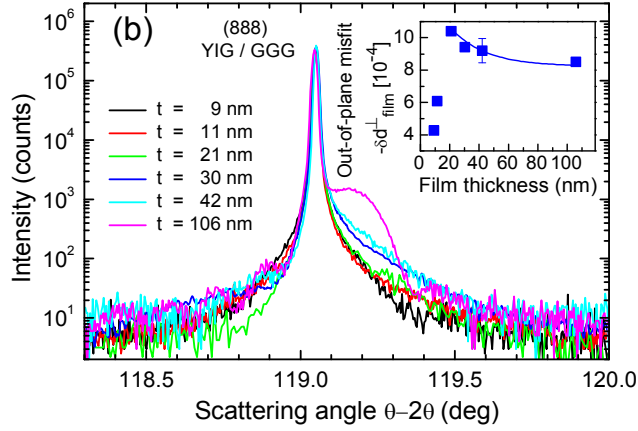

FIG. 2. (a) Combined high-resolution reciprocal space maps around the asymmetric YIG/GGG (088) Bragg reflection of a 21-nm-thin single-crystalline YIG LPE film. The inset shows the corresponding symmetric YIG/GGG (444) peak: measurements were carried out using a position-sensitive detector. (b) HR-XRD triple-axis $\theta$–$2\theta$ scans around the symmetric YIG/GGG (888) peak for various film thicknesses. The inset shows the (vertical) out-of-plane misfit vs. film thickness (the solid line is a guide to the eyes).

Figure 2(a) shows the HR-RSM of the 21 nm YIG film grown on GGG (111) substrate, measured at the asymmetric (088) reflection in steep incidence, indicating that both the film and the substrate Bragg peak positions are almost identical. Besides the nearly symmetrical intensity distribution along the [111] out-of-plane direction (i.e. the $Q_z$ axis), there is only very weak diffuse scattering close to the Bragg peak visible, pointing towards a nearly perfect crystal lattice without significant compositional strain or geometric mosaicity. In addition, no shift of in-plane (the $Q_x$ axis) film Bragg peak position with respect to the substrate is observed. This behavior indicates a fully strained pseudomorphic film growth with a perfect coherent in-plane lattice match with the GGG substrate. The pattern of the diffuse scattering observed along the $Q_x$ axis of the symmetric (444) reflection (inset in Fig. 2(a)) is very similar to the one found for a comparable GGG substrate (not shown), indicating that the defect structure of the system is mainly defined by the substrate and/or substrate surface. Within the experimental error of $\Delta Q/Q \sim 5\times10^{-6}$ nm$^{-1}$ of the high-resolution diffractometer, the same performance was found for all investigated LPE films with thicknesses below 100 nm, clearly demonstrating coherent YIG film growth without signs of film relaxation.

High-resolution triple-axis coupled $\theta$–$2\theta$ scans at the (888) and (444) symmetrical reflections (angular accuracy better than 1.5") were carried out to define the strain and film thicknesses of the YIG films. Figure 2(b) shows the results obtained at the (888) reflection. Under these conditions, the Bragg reflection of the 106 nm thick YIG layer is clearly visible as a shoulder of the (888) GGG substrate reflection at higher diffraction angles and this indicates a smaller out-of-plane value for the lattice parameter $a^{888}$ than for the GGG substrate. This is characteristic for tensely stressed "pure" YIG LPE films [50,63]. For LPE films with a thickness significantly less than 100 nm, however, only simulations can provide the structural parameters. For this reason, the diffracted signals shown in Fig. 2(b) were simulated and fitted. Using the best fit of both, the (444) and (888) reflections, the out-of-plane lattice misfit values $\delta d^{\perp}_{\text{film}} = \left(d^{\perp}_{\text{film}} - d^{\perp}_{\text{substrate}}\right)/d^{\perp}_{\text{substrate}} = -\Delta Q^{\perp}/Q^{\perp}$ were determined (see Table 1). Assuming a fully pseudomorphic [111]-oriented system, the in-plane stress of the YIG film can be calculated by $\sigma_{\parallel} = -2c_{44}\delta d^{\perp}_{\text{film}}$ (see the Supplemental Material [69] for a detailed derivation and references therein [61,70,71]). The in-plane biaxial $\varepsilon_{\parallel}$ and out-of-plane uniaxial $\varepsilon_{\perp}$ strains can be calculated as well using the stiffness tensor components $c_{11}$, $c_{12}$, and $c_{44}$ for which we use averaged values taken from [72,73] (see also Supplemental Material [69]). The resulting parameters are listed in Table I.

The inset in Fig. 2(b) shows the out-of-plane misfit as a function of the film thickness. A weak monotonous increase of $\delta d^{\perp}_{film}$ with decreasing film thickness is observed between 106 nm and 21 nm. The same behavior was reported by Ortiz *et al.* [61] for compressively strained EuIG and TbIG PLD-grown films with film thicknesses down to 4 nm and 5 nm, respectively. However, for our thinnest LPE films with $t \sim 10$ nm, the out-of-plane misfit rapidly drops. Such a significant change of the misfit with respect to the film thickness was only mentioned for considerably compressively strained YIG PLD films by O. d'Allivy Kelly *et al.* [17]. They assume, that this effect indicated a critical film thickness (below 15 nm) for strain relaxation, but did not explain it in their letter.

For semiconductor LPE films, however, it is known, that interdiffusion processes at the film/substrate interfaces can generate continuous composition profiles in the diffusion zone without abrupt changes in the lattice parameters, which lead to modified stress profiles depending on the thickness of the epilayers (see e.g. [74]). A possible explanation for the observed behavior could therefore be the presence of a smoothly changing lattice parameter value in the interface region. Such composition profiles have recently been discussed for YIG films grown on GGG substrates by high-temperature and long-time laser MBE deposition experiments [75], and transition layer thicknesses have been modeled based on polarized neutron and X-ray reflectometry techniques. The probability of the existence of such a thin continuous transition layer and its influence on the magneto-static film properties will be discussed below.

TABLE I. Structural parameters of the YIG LPE films grown on GGG (111) substrates: film thickness $t$ measured by HR-XRD, RMS roughness obtained by AFM, vertical lattice misfit $\delta d^{\perp}_{film}$ obtained by HR-XRD, in-plane strain $\varepsilon_{\parallel}$ and out-of-plane strain $\varepsilon_{\perp}$ and the resulting in-plane stress $\sigma'_{\parallel}$.

| $t$ (nm) | roughness (nm) | $\delta d^{\perp}_{film}$ $\times 10^{-4}$ | $\varepsilon_{\parallel}$ $\times 10^{-4}$ | $\varepsilon_{\perp}$ $\times 10^{-4}$ | $\sigma'_{\parallel}$ $\times 10^8$ Pa |
|---|---|---|---|---|---|
| 9 | - | -4.3 | 2.3 | -2.0 | 0.7 |
| 11 | 0.4 | -6.1 | 3.3 | -2.8 | 0.9 |
| 21 | 0.2 | -10.4 | 5.6 | -4.8 | 1.6 |
| 30 | 0.2 | -9.4 | 5.1 | -4.3 | 1.4 |
| 42 | 0.3 | -9.2 | 5.0 | -4.2 | 1.4 |
| 106 | 0.4 | -8.5 | 4.6 | -3.9 | 1.3 |
| ±1 | ±0.1 | ±0.7 | ±0.4 | ±0.4 | ±0.1 |

*2. Chemical composition studied by Rutherford Backscattering spectrometry*

Besides the epitaxial perfection, the chemical composition of the films is of interest to estimate deviations from the ideal $Y_3Fe_5O_{12}$ stoichiometry and to detect impurity elements. Therefore, RBS measurements were performed for selected LPE films. As an example, Fig. 3 shows the random spectrum of a 30 nm thick YIG film on GGG substrate. The inset presents the main part of the spectrum. Applying the NDF software, the computed curve (solid line) matches perfectly the experimental one (symbols). This enables us to determine the Fe:Y ratio. As for all investigated LPE films, the Fe:Y ratio was determined to be $R = 1.67$, which corresponds to the ideal iron garnet stoichiometry with Fe:Y = 5:3. At higher magnifications of the backscattering yield in Fig. 3, a very low intensity signal can be observed at ion energies higher than for backscattering on gadolinium atoms from the GGG substrate. Although the intensity is rather low, it can be attributed to heavy impurity elements present to a very low amount over all in the YIG film. We assign this signal to lead and platinum. These elements may come from the solvent and the crucible during the deposition of the YIG film. After background correction a total quantity of $(0.08 \pm 0.02)$ at.% for the sum of both elements could be determined. This corresponds to $0.01 < x + y < 0.02$ formula units of the nominal film composition $(Y_{3-x-y}Pb_xPt_y)(Fe_{5-x-y}Pb_xPt_y)O_{12}$. In a first approximation, for the calculation of the RBS spectra, it was assumed, that both elements contribute in equal parts to the high-energy signal. So, the calculated spectrum takes into account the existence of 0.04 at.% lead and 0.04 at.% platinum within the YIG layer. This yields a good representation of the separated signal for these two elements.

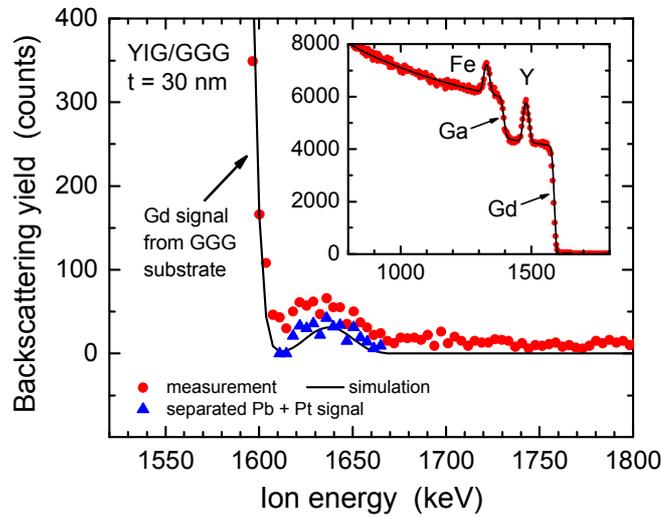

FIG. 3. Energy spectrum of 1.8 MeV He ions backscattered on the YIG/GGG sample with a YIG film thickness of $t$ = 30 nm. The inset shows the main part of the spectrum with the edges of the substrate elements Gd and Ga and the Fe and Y peak from the YIG film.

*3. Crystalline perfection studied by high-resolution transmission electron microscopy*

To analyze the film lattice perfection as well as the heteroepitaxial intergrowth behavior, HR-TEM investigations were performed. A cross-sectional image of an 11 nm thin YIG film on a GGG substrate makes it possible to visualize both, the entire YIG film volume up to the film surface and the interface in a magnified HR-TEM microscope image (see Fig. 4(a)). Besides the perfect film/substrate interface, neither structural lattice defects nor significant misalignment could be observed in the coherently strained YIG film lattice up to the film surface.
To prove the homogeneity of the bulk composition and the performance of the film/substrate interface, HAADF-STEM imaging (Fig. 4(b)) together with element mapping, based on EDXS analysis (Figs. 4(c)-(g)), were performed. The corresponding HAADF-STEM image in Fig. 4(b) allows clearly resolving the film/interface region due to the significant difference of the atomic number contrast. Because of the uniform spatial distribution of both, the film (Y, Fe, O) and the substrate elements (Gd, Ga, O), which are independently represented by different colors in Figs.

4(c)-(g), a homogeneous composition over the entire YIG film can be confirmed. Small brightness variations within the element maps (on the right hand side) result from slight thickness variations of the classically prepared TEM lamella. Neither an intermixing of the substrate nor of the film elements at the YIG/GGG interface is observed in the element maps within the EDXS detection limit, which is estimated to be slightly below 1 at.-% for the measuring conditions used. For that reason, tiny Pb and Pt contributions in the YIG film, as shown by RBS (see Fig. 3), where not detected here.

To evaluate the lateral element distributions across the film near the film/substrate interface, quantified line scans were performed as presented in Fig. 4(h). Using the 10%-to-90% edge response criterion, it shows a transition width of $(1.9 \pm 0.4)$ nm at the interface. This is lower than the observed 4-6 nm non-magnetic dead layer reported for YIG films deposited by RF magnetron sputtering [76], and the about 4 nm or the 5–7 nm deep Ga diffusion observed for PLD [77] or laser molecular beam epitaxy (MBE) [75], respectively**.** However, at some positions of the sample's cross-section we found a reduced YIG film thickness on a wavy GGG surface (not shown), which we attribute to a possible etch-back of the substrate at the beginning of film growth or an already existing wavy substrate surface. For further growth experiments, a careful characterization of the substrate surfaces by AFM should, therefore, be performed. The TEM investigations show, that the LPE technology is suitable for growing nanometer-thin YIG films without lattice defects and without significant interdiffusion at the film/substrate interface, which are necessary preconditions for undisturbed spin-wave propagation and low ferromagnetic damping losses.

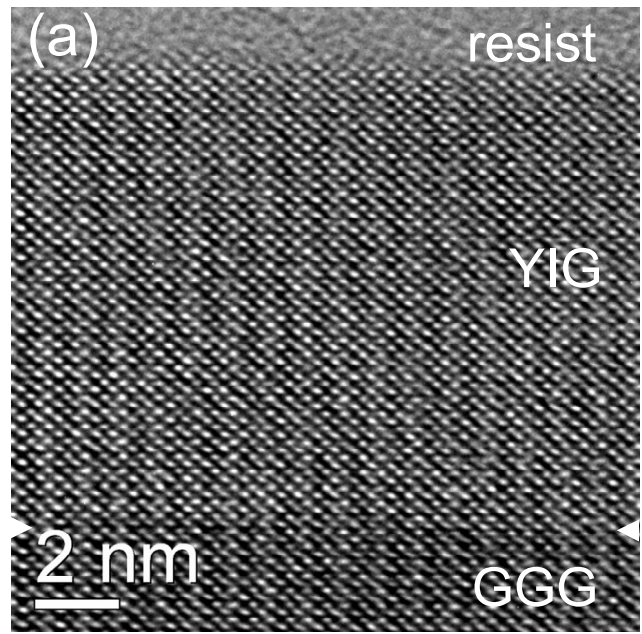

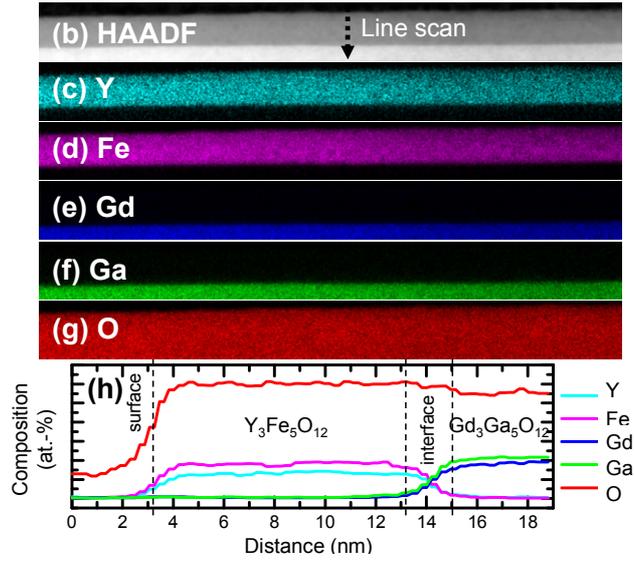

FIG. 4. (a) Cross-sectional high-resolution TEM image of the 11-nm-thin YIG/GGG (111) film. The arrows mark the YIG/GGG interface. (b) HAADF-STEM image highlighting the well-separated YIG/GGG interface. (c-g) EDXS element maps of the 11-nm-thin YIG/GGG (111) film cross-section. (h) Line scan as marked in (b) of the elemental concentrations across the film thickness.

B. Static and dynamic magnetization characterization of nanometer-thin YIG films

After gaining insight into the YIG film microstructure, we want to link these properties to the FMR performance to find out, which of them plays an essential role in the observed magneto-static and dynamic behavior. Therefore, FMR measurements were carried out within a frequency range of 1 to 40 GHz, with the external magnetic field either parallel to the surface plane of the sample along the $H \parallel [11\text{-}2]$ film direction or perpendicular to it ($H \parallel [111]$). In addition, angle-dependent measurements, i.e., varying the angle $\theta_H$ of the external magnetic field (polar angular dependence, where $\theta_H = 0$ is the sample's normal [111] direction) or the azimuth angle $\varphi_H$ (in-plane angular dependence, where $\varphi_H = 0$ is the sample's horizontal [1-10] direction), were performed at $f = 10$ GHz. These four measurement 'geometries' allow to determine Landé's $g$-factor, effective magnetization $4\pi M_{\text{eff}}$, and anisotropy fields from the resonance field dependence and to disentangle the damping contributions from the linewidth dependence [78,79].

The FMR resonance equations to fit the angle- and frequency-dependencies (see eqs. (S22), (S23) in the Supplemental Material [69] for in-plane and out-of-plane bias field conditions after Baselgia *et al.* [80]) are derived from the free energy density of a cubic (111) system [81]:

$$\begin{aligned} F = &-M_s \cdot H[\sin\theta\sin\theta_H \cos(\varphi - \varphi_H) + \cos\theta\cos\theta_H] \\ &+ (2\pi M_s^2 - K_{2\perp})\cos^2\theta - K_{2\parallel}\sin^2\theta\cos^2(\varphi - \varphi_u) \\ &+ K_4\left(\frac{1}{3}\cos^4\theta + \frac{1}{4}\sin^4\theta - \frac{\sqrt{2}}{3}\sin^3\theta\cos\theta\sin 3\varphi\right) \end{aligned}, \quad (1)$$

where $K_{2\perp}$, $K_{2\parallel}$, and $K_4$ are the uniaxial out-of plane, uniaxial in-plane, and cubic anisotropy constants, respectively. $\varphi$ and $\theta$ are the angles of the magnetization. Angle $\varphi_u$ allows for a rotation of the uniaxial anisotropy direction with respect to the cubic anisotropy direction.

*1. Frequency-dependent FMR linewidth analysis*

To investigate the influence of different contributions on the overall magnetic damping, we model the field-swept peak-to-peak linewidth $\Delta H_{pp}$ of our YIG (111) films as a sum of four contributions [79,82]:

$$\Delta H_{pp} = \Delta H_G + \Delta H_{mos} + \Delta H_0 + \Delta H_{TMS}, \qquad (2)$$

where $\Delta H_G$ is the Gilbert damping, $\Delta H_{mos}$ the mosaicity, $\Delta H_0$ the inhomogeneous broadening, and $\Delta H_{TMS}$ is the two-magnon scattering contribution, respectively. Note, that all linewidths in this paper are peak-to-peak linewidths, even if not explicitly stated.
The intrinsic Gilbert damping is given by

$$\Delta H_G = \frac{4\pi\alpha}{\sqrt{3}\gamma\Xi} f, \qquad (3)$$

where $\gamma = g\mu_B \hbar$ is the gyromagnetic ratio and $\Xi$ is the dragging function. The dragging function is a correction factor to the linewidth needed in field-swept FMR measurements if **H** and **M** are not collinear (see e.g. [82]). For **H** ∥ **M** follows $\Xi = 1$.
The inhomogeneity term $\Delta H_{mos}$ accounts for a spread (distribution) of the effective magnetization $4\pi M_{eff}$ [82,83] given by the parameter $\delta 4\pi M_{eff}$:

$$\Delta H_{mos} = \frac{2}{\sqrt{3}} \left| \frac{\partial H_{res}}{\partial 4\pi M_{eff}} \right| \delta 4\pi M_{eff}. \qquad (4)$$

$\Delta H_0$, i.e. the zero-frequency linewidth, is a general broadening term accounting for other inhomogeneities of the sample, such as the microwave power dependence of the linewidth in YIG (see, e.g., [84]) and systematic fit errors: for example, consistently narrower total full-width at half-maximum linewidths $\Delta H_{FWHM}$ of up to 0.5 Oe were determined by additional frequency-swept measurements at a microwave power of -10 dBm compared to the field-swept measurements at 0 dBm discussed here.
All kinds of inhomogeneous broadening (including $\Delta H_{mos}$) are caused by slightly different resonance fields in parts of the sample. These individual resonance lines might be still resolvable at low frequencies, where Gilbert damping is not large enough yet–especially for YIG. However, at higher frequencies, these lines become broader and eventually coalesce to a single (apparently broadened) line, which even might exhibit small shoulders or other kinds of asymmetry. Hence, what might be nicely fit with a single line at high frequencies might cause difficulties at low frequencies and sub-mT linewidths. The effect on fitting the anisotropy constants from the resonance fields is not so sensitive. If the resonance lines cannot be disentangled or the line is not entirely Lorentzian-shaped anymore, the fit might overestimate the true linewidth resulting in a systematic broader line accounted for by $\Delta H_0$.

The last term in Eq. (2), $\Delta H_{TMS}$, covers the two-magnon scattering contribution, which is an extrinsic damping mechanism due to randomly distributed defects. For the in-plane frequency-dependence it reads [78,79,82,85-87]:

$$\Delta H_{TMS} = \frac{2}{\Xi\sqrt{3}} \Gamma \cdot \sin^{-1} \sqrt{\frac{\sqrt{f^2 + \left(\frac{f_0}{2}\right)^2} - \frac{f_0}{2}}{\sqrt{f^2 + \left(\frac{f_0}{2}\right)^2} + \frac{f_0}{2}}}, \qquad (5)$$

where $f_0 = \gamma 4\pi M_{eff}$ and $\Gamma$ is the two-magnon scattering strength.

Each of the contributions has a characteristic angle and frequency dependence. Overall, the linewidth vs. frequency dependencies and the linewidth vs. angle dependencies can be described with one set of parameters.

As we will see, the applied model fits very well to the experimental results and allows for disentangling the contributions that are responsible for the frequency dependence of the linewidth. At first, we discuss the different damping contributions. Then, we go into details for the individual magneto-static parameters, the relevant anisotropy contributions mentioned above, which provided also the base input for the fit parameters for the frequency-dependent FMR linewidth of our YIG films.

In Figure 5, the obtained frequency-dependent peak-to-peak linewidths $\Delta H_{pp}$ (symbols) for the four thicknesses 11 nm, 21 nm, 30 nm, and 42 nm are presented. The red (solid) curves represent fits using Eq. (2). Figure 5(a) shows data and fits for the out-of-plane bias field configuration ($\theta_H = 0°$) and Fig. 5(b) for field-in-plane ($\theta_H = 90°$), respectively. As mentioned above, due to a quite complex shape of the resonance lines below ~15 GHz for $\theta_H = 0°$ (with more absorption lines needed to reflect the shape of the spectrum than for higher frequencies) the linewidths could not anymore be evaluated unambiguously with the required precision for films with thicknesses above 11 nm. However, for the thinnest film, the evaluation was possible and the overall fit exhibits a linear behavior down to 1 GHz. This means, in the field-out-of-plane geometry, the main contribution to the damping is the Gilbert damping α, which can be determined from the linear slope according to Eq. (3). As it is known from two-magnon scattering (TMS) theory [85,86], there is no TMS contribution if $M$ is perpendicular to the sample plane. The only remaining contribution is the inhomogeneous broadening given by the zero-frequency offset

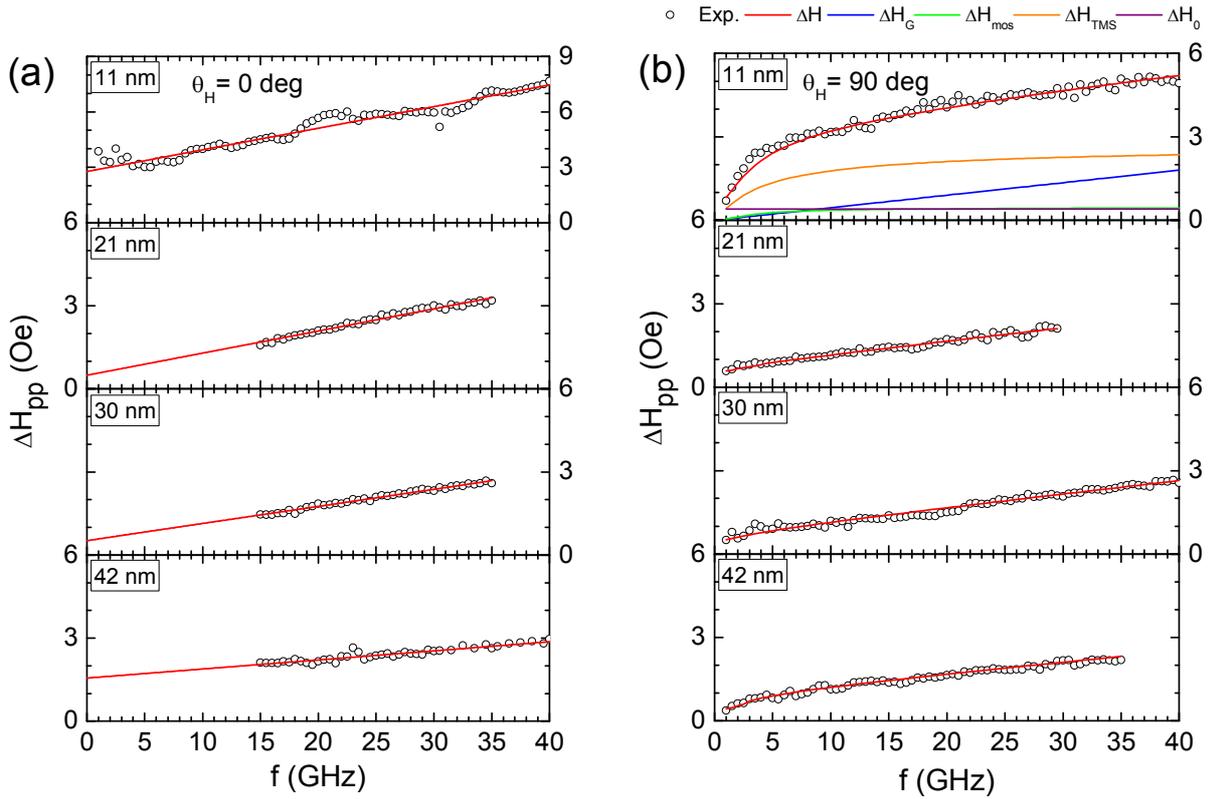

FIG. 5: Frequency dependence of the linewidth with magnetic field (a) perpendicular-to-plane and (b) in-plane. The red (solid) lines are fits to the data. For the 11-nm sample the individual contributions to the total linewidth are shown in the top-right panel. Note the different *y*-axis scaling for the 11 nm sample in the top-left panel.

From these out-of-plane measurements, the Gilbert damping coefficients could be determined, ranging from $\alpha = 0.9 \times 10^{-4}$ for the 42-nm-thick sample to $\alpha = 2.0 \times 10^{-4}$ for the 21 nm sample, which is about twice the value obtained from in-plane measurements (as discussed below). For the ultrathin 11 nm film, a slightly increased Gilbert damping coefficient of $\alpha = 2.7 \times 10^{-4}$ and a significantly enlarged zero-frequency linewidth of 2.8 Oe were found. As mentioned above, the reason for the larger offset might be an apparent unresolvable broadening due to inhomogeneity. For the 21 and 30 nm sample, the zero-frequency intercept is about $\Delta H_0 = 0.5$ Oe, in contrast to $\Delta H_0 = 1.5$ Oe for the 42 nm sample. This indicates, that the 42 nm sample, in contrast to the thinner samples, seems to have additional microstructural defects, leading to a superposition of lines. This is very likely, because the inhomogeneous broadening previously reported for 100 nm YIG LPE films was also in the range of $\Delta H_0 = 0.5$-$0.7$ Oe [50].

In Fig. 5(b), the results of the corresponding in-plane field configuration are given. For the 11 nm sample, the four individual fit contributions considered in the fit according to Eq. (2) are depicted by solid curves. This sample shows a significant curvature. The 42 nm sample also shows a small curvature, whereas the other two samples only have a weak curvature at lower frequencies. This curvature usually hints to a contribution from two-magnon scattering, but can also be due to a spread of the effective magnetization. Note, the frequency-dependence of the mosaicity and TMS term look quite similar at higher frequencies, but show a different curvature at lower frequencies. Hence, the shape of the curve and, thus, the fit reveal, that it is due to a spread of the effective magnetization, $\delta 4\pi M_{\text{eff}}$ as given by Eq. (4), which lies in the range of 0.4 to 0.9 G. For the 11 nm sample, this value is larger, i.e., $\delta 4\pi M_{\text{eff}}$ = 3.2 G, and in addition one needs a small TMS damping contribution of $\Gamma$ = $1.5 \times 10^7$ Hz for a proper fit (see Table II). This is again a distinctive sign, that the 11 nm sample has significantly different structural and/or magnetic properties, leading to the additional linewidth contributions. The Gilbert damping coefficients of all four samples in in-plane configuration are $\alpha \leq 1.3 \times 10^{-4}$ and correspond to the best values reported earlier for 100 nm YIG LPE films [50]. These are also lower than for a recently reported 18 nm YIG LPE film [51]. Thus, at room temperature, no significant increase in Gilbert damping could be observed for LPE films down to 10 nm with decreasing thickness. This contrasts with various references for PLD and RF-sputtered YIG films grown on (111) GGG substrates [88-92].

TABLE II. Magnetic damping parameters of the LPE (111) YIG films: film thickness $t$, in-plane Gilbert damping parameter $\alpha_\parallel$, inhomogeneous broadening $\Delta H_{0\parallel}$, spread of effective magnetization $\delta 4\pi M_{\text{eff}}$ and two-magnon scattering contribution $\Gamma$.

| $t$ (nm) | $\alpha_\parallel$ ($\times 10^{-4}$) | $\Delta H_{0\parallel}$ (G) | $\delta 4\pi M_{\text{eff}}$ (Oe) | $\Gamma$ ($10^7$ Hz) |
|---|---|---|---|---|
| 11 | 1.2 | 0.4 | 3.2 | 1.5 |
| 21 | 1.3 | 0.6 | 0.4 | 0 |
| 30 | 1.2 | 0.4 | 0.7 | 0 |
| 42 | 1.0 | 0.4 | 0.9 | 0 |
| accuracy | ±0.2 | ±0.2 | ±0.3 | ±0.3 |

All field-in-plane linewidth parameters of the investigated samples are summarized in Table II. It is obvious, that inhomogeneous contributions, i.e., those originating from magnetic mosaicity $\delta 4\pi M_{\text{eff}}$, are very small for the samples without two-magnon scattering. This confirms the high microstructural perfection and homogeneity of the volume and interfaces of the LPE-grown films with film thicknesses larger than 11 nm. Contributions to two-magnon scattering appear to occur only for LPE films with a thickness of less than 21 nm thick.

*2. Analysis of magnetic anisotropy contributions*

In the following, we will discuss the anisotropy contributions, which provided the base input for the fit parameters used for the frequency-dependent FMR linewidth curves shown above. All curves were fitted iteratively with the respective resonance equation (see Eqs. (S22) and (S23) in the Supplemental Material [69]) to retrieve a coherent set of fit parameters. The fit parameters are listed in table III. Since the saturation magnetization and the in-plane stress are known from VSM measurements and HR-XRD investigations, the anisotropy constants $K$ can be calculated from the anisotropy fields determined by FMR.

TABLE III. Magneto-static parameters of the YIG LPE films of thickness $t$: Landé's $g$-factor, effective magnetization $4\pi M_{\text{eff}}^{\text{exp}}$, cubic anisotropy field $2K_4/M_s$, and uniaxial in-plane anisotropy field $2K_{2\parallel}/M_s$ determined from FMR, saturation magnetization $4\pi M_s$ determined from VSM, stress-induced anisotropy field $2K_\sigma/M_s$ calculated from X-ray diffraction data, resulting out-of-plane uniaxial anisotropy field $2K_{2\perp}/M_s$ and effective magnetization $4\pi M_{\text{eff}}^{\text{cal}}$, cubic anisotropy constant $K_4$, stress-induced anisotropy constant $K_\sigma$, and out-of-plane uniaxial anisotropy constant $K_{2\perp}$.

| $t$ (nm) | $g$ | $4\pi M_{\text{eff}}^{\text{exp}}$ (G) | $2K_4/M_s$ (Oe) | $2K_{2\parallel}/M_s$ (Oe) | $4\pi M_s$ (G) |
|---|---|---|---|---|---|
| 11 | 2.015 | 1566 | -93 | 2.0 | 1494 |
| 21 | 2.016 | 1647 | -79 | 0.8 | 1819 |
| 30 | 2.015 | 1677 | -79 | 0.6 | 1830 |
| 42 | 2.014 | 1699 | -86 | 1.1 | 1860 |
| accuracy | ±0.002 | ±13 | ±2 | ±3 | ±41 |

| $t$ (nm) | $2K_\sigma/M_s$ (G) | $2K_{2\perp}/M_s$ (G) | $4\pi M_{\text{eff}}^{\text{cal}}$ (G) | $K_4$ ($10^3$ erg/cm$^3$) | $K_\sigma$ ($10^3$ erg/cm$^3$) | $K_{2\perp}$ ($10^3$ erg/cm$^3$) |
|---|---|---|---|---|---|---|
| 11 | 65 | 127 | 1368 | -5.5 | 3.9 | 7.5 |
| 21 | 91 | 143 | 1676 | -5.7 | 6.6 | 10.4 |
| 30 | 82 | 135 | 1696 | -5.8 | 6.0 | 9.8 |
| 42 | 79 | 136 | 1724 | -6.4 | 5.8 | 10.1 |
| accuracy | ±7 | ±4 | ±40 | ±0.3 | ±0.4 | ±0.5 |

The $g$-factor of the samples was determined from the frequency dependencies of the resonance field. There was no significant thickness dependence observed yielding a value of $g = 2.015(1)$ for all samples. The cubic anisotropy field $2K_4/M_s$ was found to be nearly constant, and the average value is -84(2) Oe, which is in good agreement to reported values of -85 Oe for a 120 micrometer thick LPE film [81] and of about -80 Oe for a 18 nm thin LPE film [51]. Our calculated anisotropy constants $K_4$ are almost always in the range between -5.7×10$^3$ and -6.4×10$^3$ erg/cm$^3$, which corresponds to YIG single crystal bulk values at 295 K [93]. Furthermore, a rather weak in-plane uniaxial anisotropy field $2K_{2\parallel}/M_s$ of about 0.6–2 Oe was found, which had already been determined for 100 nm YIG LPE films [50].

The stress-induced anisotropy constant $K_\sigma$ and anisotropy field $2K_\sigma/M_s$ are calculated according to Ref. [94] (for details, see Eqs. (S14), (S15), (S18) in the Supplemental Material [69]). $2K_\sigma/M_s$ is small and in the same order of magnitude as the cubic anisotropy field $2K_4/M_s$, but with opposite sign. Due to the observed monotonous increase of the out-of-plane lattice misfit (see inset in Fig. 2(b)), $2K_\sigma/M_s$ grows with decreasing film thickness until it declines significantly at a film thickness below 21 nm. However, the observed stress values are almost an order of magnitude smaller than, e.g., for as-deposited YIG PLD films on GGG (111) under compressive strain (see, e.g., Refs. [17,23,35,36]). Only by a complex procedure, applying mid-temperature deposition, cooling, and post-annealing treatment, authors of Ref. [95] succeeded in a change from compressively to tensely strained YIG films. These samples then exhibited the same stress-induced anisotropy constant as it was observed for our YIG LPE films.

In the following, we take a closer look to the contributions to the out-of-plane uniaxial anisotropy field $H_{2\perp} = 2K_{2\perp}/M_s$. A general description for magnetic garnets has been given for example by Hansen [94]. Applied to thick [43,64] as well as to thin epitaxial iron garnet films (see, e.g., [37,56,59,60,62]), the out-of-plane uniaxial anisotropy field $H_{2\perp}$ is mainly determined by the magnetocrystalline and uniaxial anisotropy contributions. While the former refers to the direction of magnetization to preferred crystallographic directions in the cubic garnet lattice, the latter originates from lattice strain and growth conditions. Due to the very low supercooling (≤5K), growth-induced contributions, usually observed for micrometer YIG films with larger Pb impurity contents, can be neglected in the case of our nanometer-thin YIG LPE films (see e.g. [64]). Thus, $H_{2\perp}$ can be determined quantitatively by summing the cubic magnetocrystalline anisotropy (first term, determined by FMR) and the stress-induced anisotropy (second term, determined by XRD),

$$H_{2\perp} = -\frac{4}{3}\frac{K_4}{M_s} + \frac{2K_\sigma}{M_s}, \tag{6}$$

or expressed for the (111) substrate orientation (see also SM [69] and Ref. [93,96]) by:

$$H_{2\perp} = \frac{-4K_4 - 9\sigma'_\parallel \lambda_{111}}{3M_s}. \tag{7}$$

Using the experimentally determined first-order cubic anisotropy constant $K_4$ and the in-plane stress component $\sigma'_\parallel$ from Tables I and III along with the room-temperature magnetostriction coefficient $\lambda_{111}$ [94], the uniaxial anisotropy field $H_{2\perp}$ can be calculated, if the saturation magnetization $M_s$ is known. $M_s$ can be obtained with appropriate accuracy for example from VSM or SQUID measurements, if the sample volume is exactly known.
Magnetic hysteresis loops of YIG LPE films recorded at room-temperature by VSM measurements with in-plane applied magnetic field are shown in Fig. 6. The paramagnetic contribution of the GGG substrate was subtracted as described in Ref. [50]. Extremely small coercivity fields with $H_c$ values of ~ 0.2 Oe were obtained for all YIG/GGG samples with the exception of the 21 nm film. These values are comparable with the best gas phase epitaxial films [17,39,76], but the measured saturation fields with $H_s < 2.0$ Oe are significantly smaller. All films exhibit nearly in-plane magnetization due to the dominant contribution of form anisotropy. Apart from the thinnest sample, the saturation moments determined are not thickness-dependent (see Table III and Fig. 6) and are very close to YIG volume values determined for YIG single crystals at room temperature ($4\pi M_s$ ~ 1800 G) [93,97]. However, the observed decrease of the saturation magnetization in such films with a thickness of about 10 nm is significant and will be discussed below.

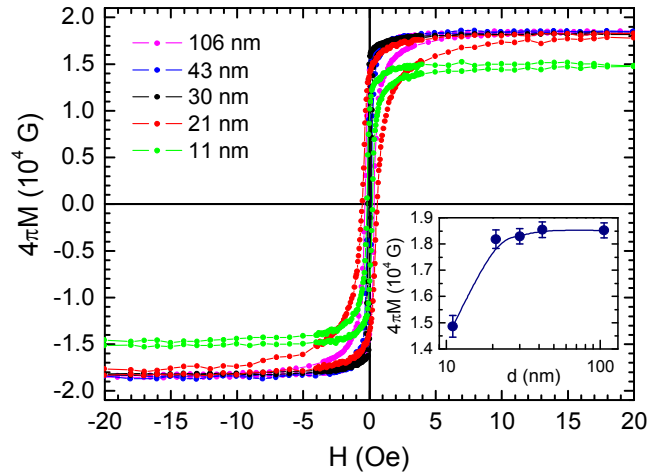

FIG.6: Magnetization loops $M(H)$ of YIG films at room temperature as a function of the in-plane magnetic field. The inset shows the thickness-dependent saturation magnetization (the solid line is a guide to the eyes).

However, for nanometer-thin films, it is a big challenge to determine $M_s$ precisely enough, because too large errors can arise from the film's volume calculation. While the surface area of the sample can be determined with sufficient precision by optical microscopy, thickness measurements with X-ray or ellipsometry methods can lead to thickness errors in the range of ±1 nm due to very small macroscopic morphology or roughness fluctuations. Therefore, for films with thicknesses below 20 nm, for example, uncertainties up to a maximum of 10 percent must be considered. This could significantly affect the effective magnetization $4\pi M_{eff}$, which can be calculated based on the measured $M_s$ values by

$$4\pi M_{eff} = 4\pi M_s - H_{2\perp}. \qquad (8)$$

This fact can explain the large difference between the calculated $4\pi M_{eff}^{cal}$ and the measured $4\pi M_{eff}^{exp}$ values for the 11 nm thin film discussed below, while a much better agreement was achieved for the thicker films (see Table III).

As expected from micrometer-thick YIG LPE films grown on GGG (111) substrates [81], the out-of-plane uniaxial anisotropy field $H_{2\perp}$ and the out-of-plane uniaxial anisotropy constants $K_{2\perp}$ show, that completely pseudomorphically strained, nanometer-thin LPE films exhibit no pronounced magnetic anisotropy. Small changes of the in-plane stress $\sigma'_{\parallel}$ (see Table I) and thus also in the stress-induced anisotropy $2K_\sigma/M_s$ (or $K_\sigma$) have no significant influence on the out-of-plane uniaxial anisotropy $H_{2\perp}$ (see Table III). A comparable $H_{2\perp}$ value is also expected for films thicker than 42 nm, since the out-of-plane lattice misfit $\delta d^\perp_{film}$ tends to a constant value (see inset in Fig. 2 (b)). This is in contrast to Ref. [51], where the uniaxial anisotropy field of YIG LPE films becomes negative above a film thickness of about 50 nm.

### 3. Thickness-dependent analysis of the effective magnetization field

To verify the trend of the calculated $4\pi M_{eff}^{cal}$ values for decreasing film thicknesses, one can compare the effective magnetization with the experimentally determined one. This was done for 18 YIG films with thicknesses ranging from 10 to 120 nm, including the four samples from above. All films were grown during the same run under nearly identical conditions. Only the growth temperature was varied within a range of 5 K. This time, the FMR was measured with a constant external magnetic field applied in-plane and sweeping the frequency.

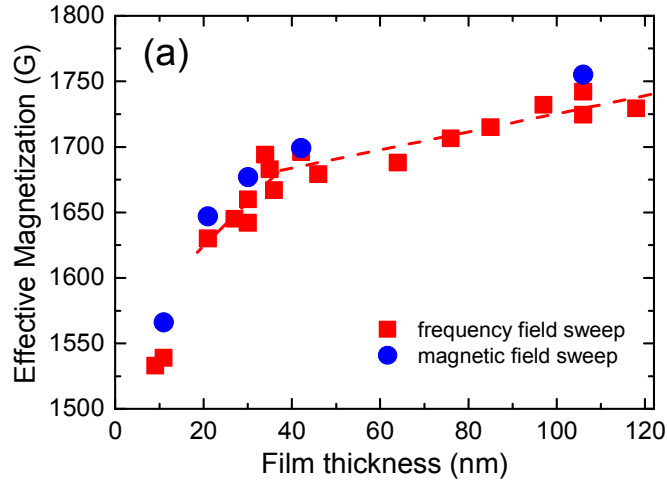

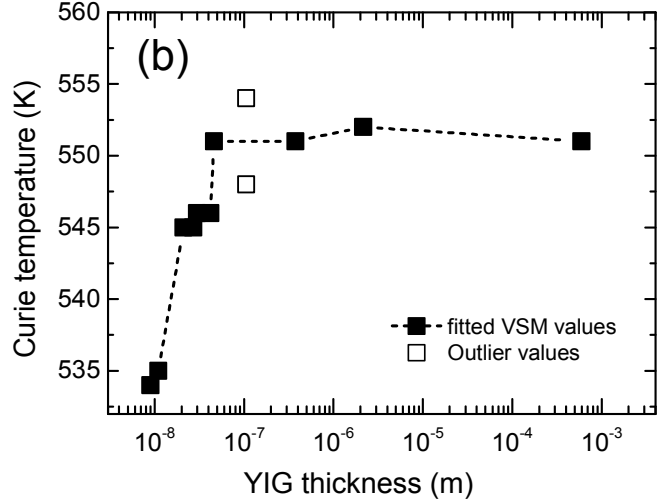

Fig. 7. (a) Thickness dependence of the effective magnetization $4\pi M_{\text{eff}}$. Blue circles denote measurements taken by field-sweep and red squares denote frequency-swept measurements, respectively. (b) Thickness dependence of the Curie temperature $T_c$. The dashed line is a guide to the eyes.

In Fig. 7(a), the obtained thickness dependence of the effective magnetization $4\pi M_{\text{eff}}$ (squares) is presented and a monotonous decrease of $4\pi M_{\text{eff}}$ with film thickness reduction can be observed. Below 40 nm, the slope of the curve increases and, for the thinnest films, there is a significant drop of about 100 G. This behavior has been confirmed by in-plane FMR magnetic field-sweep measurements for selected samples (circles), as discussed before. The values are listed in Table III. Similar results have been reported for YIG PLD films by Kumar *et al.* [53].

If we compare the experimental values with the calculated ones in Table III, then the same trend of a steady reduction of the effective saturation magnetization with decreasing film thickness can be observed. The deviation between both $4\pi M_{\text{eff}}$ values is approximately 1–2 %, except for the 11 nm film. Hence, the saturation magnetization used to calculate the effective saturation (according to equation (8)) does not appear to be as error-prone as it could be due to an inaccuracy in the film thickness determination. Therefore, we speculate, that the significant drop of $4\pi M_{\text{eff}}$ for the 11 nm thin film can be explained by the observed reduction of $4\pi M_s$ (see Table III).

A similar behavior for $4\pi M_s$ was reported for thin PLD or magnetron-sputtered YIG films, and different explanations were given [76,56,91]. One reason for a reduced saturation magnetization could be an intermixing of substrate and film elements at the GGG/YIG interface, whereby a gradual change of the film composition is assumed [75,77]. In particular, gallium ion diffusion into the first YIG atomic layers will lead to magnetically diluted ferrimagnetic layers at the interface, due to the fact, that magnetic Fe ions are replaced by diamagnetic Ga ions in the various magnetic sublattices. This assumption is supported by recent reports of YIG films on GGG substrates. One reports about a 5–7 nm deep Ga penetration found in laser-MBE films [75]. Another group found a Ga penetration throughout a 13-nm-thin PLD film [98]. In these cases, high-temperature film growth above 850°C or prolonged post-annealing at temperatures of 850°C could promote such diffusion processes. In contrast, the deposition time during which the LPE samples were exposed to high temperatures above 860°C was only 5 minutes. Though, the assumed Ga diffusion depth in our YIG films should not exceed more than 2 nm according to the EDXS element maps in Fig. 4(h). In addition, the Gd diffusion in YIG films, as discussed for RF-magnetron sputtered [76,99] or PLD films [98], could lead to the incorporation of paramagnetic ions into the diamagnetic rare earth sublattice sites, which would also alter the magnetization [98]. However, no extended interdiffusion layer was observed at the film/substrate interface for our LPE films, so that the presence of a 'separate, abrupt' gadolinium iron garnet interface layer, as reported by Ref. [98], is not expected. Therefore, due to possible interdiffusion effects at temperatures of about 860°C, a gradual reduction of $M_s$ at a postulated interface layer could be the reason for the observed low saturation value for the thinnest LPE film, listed in Tab. III.

To further rule out a discrete magnetic dead layer, Curie temperature ($T_c$) measurements were performed by VSM. It is known from literature, that $T_c$ remains constant up to a film thickness of approximately four YIG unit cells [100], i.e. 2.8 nm, since one YIG unit cell length along the [111] direction amounts to $d_{111} \sim 0.7$ nm. Accordingly, the $T_c$ of "pure" YIG films with abrupt interfaces and a film thickness of ~10 nm should be equal to that of bulk material. In order to check this, temperature-dependent VSM measurements (see Fig. 7(b)) were carried out for our LPE films as well as for a bulk YIG single crystal slice, which was used as a reference. We found almost constant values of $T_c = (551\pm2)$ K for sample thicknesses between 46 nm (thin film) and 0.55 mm (bulk). This is in good agreement with the literature, in which a $T_c$ of ~550 K has been reported, e.g. for a 100-nm-thin sputtered YIG film [76], while 559 K has been reported for YIG single crystals [97]. However, for our about 10-nm-thin YIG films, $T_c$ decreased significantly to ~534±1 K (Fig 7(b)), which is consistent with the observed reduction of $4\pi M_s$ listed in Table III.

Hence, the most likely explanation for the observed reduction of $4\pi M_s$ is that the YIG layers at the substrate/film interface exhibit a reduced saturation magnetization due to a magnetically diluted iron sublattice, resulting from high-temperature diffusion of gallium ions from the GGG substrate into the YIG film. While nearly zero gallium content at the film surface leads to a bulk-like value of $4\pi M_s \sim 1800$ G [93], an increased content of gallium at the film/substrate interface should, therefore, result in significantly reduced $4\pi M_s$ values. In this case, the average saturation magnetization for the entire film volume should be reduced and that could explain the observed decrease in $4\pi M_s$ to about 1500 G for the 11 nm thin LPE film. For thicker films, however, the influence of thin gallium-enriched interface layers on the entire film magnetization decreases, which explains the fast achievement of a constant Curie temperature, and thus, a constant $M_s$ with increasing thickness of the YIG volume. In order to confirm our assumptions, additional analyses, such as detailed secondary ion mass spectroscopy (SIMS) investigations, are necessary which, however, go beyond the scope of this report.

## IV. CONCLUSIONS AND OUTLOOK

In summary, we have demonstrated that LPE can be used to fabricate sub-40 nm YIG films with high microstructural perfection, smooth surfaces and sharp interfaces as well as excellent microwave properties down to a minimum film thickness of 11 nm. All LPE films with ≥21 nm thickness exhibit extremely narrow FMR linewidths of $\Delta H_{pp} <1.5$ Oe at 15 GHz and very low magnetic damping coefficients of $\alpha \leq 1.3 \times 10^{-4}$ which are the lowest values reported within an extended frequency range of 1 to 40 GHz. We were able to show that LPE-grown YIG films down to a thickness of 21 nm have the same magnetization dynamics influenced by small cubic and stress-induced anisotropy fields. The deviating magnetization dynamics of ultrathin LPE films with thicknesses of ~10 nm are probably caused by an increased inhomogeneous damping and by small two-magnon scattering contributions, and we speculate that possible inhomogeneities of the composition in the vicinity of the film/substrate interface might be the reason for this. Therefore, in further studies we will address detailed investigations of the composition of the film/substrate interface by high-resolution SIMS measurements and advanced STEM analyses to confirm a gradual change of the LPE film composition at the interface.

The results presented here encourage us to take the next step towards nano- and microscaled magnonic structures, such as directional couplers, logic gates, transistors etc. for a next-generation of computing circuits. The development of nanoscopic YIG waveguides and nanostructures is already underway and the first circuits are currently being fabricated [10,12,29]. With its scalability to large wafer diameters of up to 3 and 4 inches, LPE technology opens up an alternative way for efficient circuit manufacturing for a future YIG planar technology on a wafer scale.

## ACKNOWLEDGMENTS

We thank P. Landeros and R. Gallardo for fruitful discussions and A. Khudorozhkov for his help during the measurements. C. D. and O. S. thank R. Köcher for AFM measurements, A. Hartmann for the DSC measurements and R. Meyer and B. Wenzel for technical support. J. G. thanks A. Scholz for the support during the XRD measurements. We would like to thank Romy Aniol for the TEM specimen preparation. The use of HZDR's Ion Beam Center TEM facilities and the funding of TEM Talos by the German Federal Ministry of Education of Research (BMBF), Grant No. 03SF0451 in the framework of HEMCP are gratefully acknowledged.

This research was financially supported by the Deutsche Forschungsgemeinschaft (DFG), via Grant No. DU 1427/2-1.

# Supplemental Material:
# Low damping and microstructural perfection of sub-40nm-thin yttrium iron garnet films grown by liquid phase epitaxy

Carsten Dubs,[1] Oleksii Surzhenko,[1] Ronny Thomas,[2] Julia Osten,[2] Tobias Schneider,[2] Kilian Lenz,[2] Jörg Grenzer,[2] René Hübner,[2] Elke Wendler[3]

[1] INNOVENT e.V. Technologieentwicklung, Prüssingstr. 27B, 07745 Jena, Germany
[2] Institute of Ion Beam Physics and Materials Research, Helmholtz-Zentrum Dresden-Rossendorf, Bautzner Landstr. 400, 01328 Dresden, Germany
[3] Institut für Festkörperphysik, Friedrich-Schiller-Universität Jena, Helmholtzweg 3, 07743 Jena, Germany

## I. STRAIN CALCULATIONS

In the following we will derive the in-plane (horizontal) stress $\sigma_\parallel$ as a function of the out-of-plane (vertical) lattice misfit $\delta d_{\text{film}}^\perp$ for a [111] oriented cubic system. These calculations are based on the elasticity theory following mainly Hinckley [70], Sander [71] and Ortiz *et al.* [61]. Assuming a fully pseudomorphic system there is only one parameter to be determined: The out-of-plane lattice misfit $\delta d_{\text{film}}^\perp$. This value can be directly obtained from the data of the HR-XRD measurements and/or from the corresponding simulations of the symmetrical (444) and (888) reflections.

The vertical and parallel lattice misfits can be calculated by

$$\delta d_{\text{film}}^\perp = \frac{d_{\text{film}}^\perp - d_{\text{substrate}}^\perp}{d_{\text{substrate}}^\perp} \qquad \delta d_{\text{film}}^\parallel = \frac{d_{\text{film}}^\parallel - d_{\text{substrate}}^\parallel}{d_{\text{substrate}}^\parallel} = 0, \qquad (S1)$$

where $d^k_i$ with $i$ = [substrate, (pseudomorph) film or (cubic) relaxed film] is the (measured) net plane distances for the $k = \perp$ (vertical) or $\parallel$ (parallel) direction with respect to the substrate surface. Ignoring any dynamical diffraction effects, the out-of-plane lattice misfit can be directly determined from the measurement as follows:

$$\delta d_{\text{film}}^\perp = -\delta q_{\text{film}}^\perp = -\frac{q_{\text{film}}^\perp - q_{\text{substrate}}^\perp}{q_{\text{substrate}}^\perp} = -\frac{\Delta q^\perp}{q_{\text{substrate}}^\perp} = -\frac{\Delta\theta}{\tan\theta_B}, \qquad (S2)$$

where $q^\perp{}_{\text{film}}$ and $q^\perp{}_{\text{substrate}}$ are the derived peak positions in the Q-space of the thin film and the substrate, respectively. $\Delta\theta_B$ is the (kinematical) Bragg angular difference "thin film – substrate" and $\theta_B$ is the Bragg Peak position of the substrate, respectively. These formulas follow directly from the derivative of Bragg's law.

The in- and out-of-plane strains are given as follows:

$$\varepsilon^\perp = \frac{d_{\text{film}}^\perp - d_{\text{relaxed film}}^\perp}{d_{\text{relaxed film}}^\perp} \qquad \varepsilon^\parallel = \frac{d_{\text{film}}^\parallel - d_{\text{relaxed film}}^\parallel}{d_{\text{relaxed film}}^\parallel}. \qquad (S3)$$

The general relationship between stress and strain is defined as follows:

$$\sigma_{ij} = \mathbf{C}_{ijkl}\varepsilon_{kl}, \qquad (S4)$$

where $\sigma_{ij}$ are the stress, $\varepsilon_{kl}$ the strain and $\mathbf{C}_{ijkl}$ the second order stiffness tensors and the summation is done over the repeated indices. The subscripts $ij$ and $kl$ refer to the axes of the coordinates system of the unit cell (1,2,3 = x,y,z). The samples under investigations have a [111] out-of-plane orientation; therefore, the corresponding rotation matrices have to be applied:

$$\mathbf{C}'_{\alpha\beta\gamma\delta} = \mathbf{U}_{\alpha i}\mathbf{U}_{\beta j}\mathbf{U}_{\gamma k}\mathbf{U}_{\delta l}\mathbf{C}_{ijkl}. \quad (S5)$$

For the [111] oriented surfaces $U^{111}$ yields to

$$\mathbf{U}^{111} = \begin{pmatrix} \frac{1}{\sqrt{6}} & \frac{-1}{\sqrt{2}} & \frac{1}{\sqrt{3}} \\ \frac{1}{\sqrt{6}} & \frac{1}{\sqrt{2}} & \frac{1}{\sqrt{3}} \\ \frac{\sqrt{2}}{\sqrt{3}} & 0 & \frac{1}{\sqrt{3}} \end{pmatrix}. \quad (S6)$$

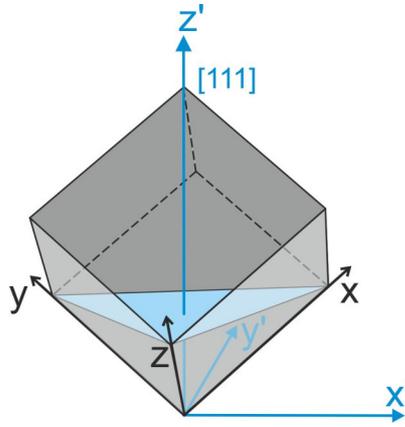

**Figure S1**: Representation of the cubic unprimed and the rotated, primed coordinate system for an <111> oriented thin film; where $x'$, $y'$ correspond to the in-plane (∥) directions and $z'$ to the out-of-plane (⊥) direction; after [71].

The in-plane stress $\sigma'_\parallel$ can be expressed in terms of the out-of-plane lattice misfit $\delta d^\perp_{film}$ obtained by XRD measurements.

For a cubic pseudomorphic system we can write:

$$\sigma_{11} = c_{11}\varepsilon_{11} + c_{12}\varepsilon_{12} + c_{13}\varepsilon_{13} \quad (S7)$$

$$\sigma'_\parallel = (c'_{11} + c'_{12})\varepsilon_\parallel + c'_{12}\varepsilon_\perp \quad (S8)$$

with

$$\varepsilon_\parallel = -\nu'\varepsilon_\perp \quad (S9)$$

resulting in:

$$\sigma'_\parallel = 6c_{44}\frac{c_{11} + 2c_{12}}{c_{11} + 2c_{12} + 4c_{44}}\varepsilon_\parallel. \quad (S10)$$

Taking the corresponding rotation matrices and relationships into account [101]:

$$\varepsilon^{\perp} = \frac{1}{1+v^{111}}\delta d^{\perp}_{film} \quad \text{and} \quad \varepsilon^{\parallel} = -\frac{v^{111}}{1+v^{111}}\delta d^{\perp}_{film}, \quad \text{(S11)}$$

where

$$v^{111} = \frac{c_{11} + 2c_{12} + 4c_{44}}{2c_{11} + 4c_{12} - 4c_{44}}. \quad \text{(S12)}$$

The in-plane stress can be now expressed in terms of the out-of-plane lattice misfit by:

$$\sigma'_{\parallel} = -2c_{44}\delta d^{\perp}_{film}. \quad \text{(S13)}$$

Here, $c_{44}$ is the component from the stiffness tensor and $\delta d^{\perp}_{film}$ is the out-of-plane lattice misfit as defined above.

## II. ANISOTROPY CALCULATIONS FOR (111) ORIENTED EPITAXIAL GARNET FILMS

The stress-induced anisotropy parameter for the cubic (111) orientation can be calculated according to [94] by

$$K_{\sigma} = -\frac{3}{2}\sigma'_{\parallel}\lambda_{111}, \quad \text{(S14)}$$

where $\sigma'_{\parallel}$ is the above calculated in-plane stress for {111} oriented thin films, and $\lambda_{111}$ is the corresponding magnetostriction constant.
The stress-induced anisotropy parameter is therefore given by:

$$K_{\sigma} = 3c_{44}\delta d^{\perp}_{film}\lambda_{111}. \quad \text{(S15)}$$

The perpendicular magnetic anisotropy field can be calculated according to [43]:

$$H_{2\perp} = H_{cub} + H_{stress} + H_{growth}. \quad \text{(S16)}$$

Assuming negligible growth-induced contributions $H_{growth}$ and applying the cubic anisotropy field for (111) film orientation obtained by FMR measurements

$$H_{cub} = -\frac{4}{3}\frac{K_4}{M_s}, \quad \text{(S17)}$$

and taking into account the stress-induced anisotropy field

$$H_{stress} = \frac{2K_{\sigma}}{M_s} = -\frac{3\sigma'_{\parallel}\lambda_{111}}{M_s}, \quad \text{(S18)}$$

the effective perpendicular anisotropy field results in

$$H_{2\perp} = -\frac{4K_4 + 9\sigma'_{\parallel}\lambda_{111}}{3M_s}. \quad \text{(S19)}$$

From the resonance conditions for the perpendicular (M ∥ [111]) magnetized epitaxial thin film, the effective saturation magnetization can be obtained by [64]

$$\frac{\omega}{f} = \left[ H_{eff} - 4\pi M_s - \frac{4}{3}\frac{K_4}{M_s} + \frac{2K_{2\perp}}{M_s} \right], \tag{S20}$$

from which the effective saturation magnetization can be calculated by

$$H_{eff} = 4\pi M_{eff} = 4\pi M_s - H_{2\perp}. \tag{S21}$$

## III. FERROMAGNETIC RESONANCE

From the free energy density given by Eq. (1) of the main text, the resonance equations have been calculated applying the approach of Baselgia *et al.* [80]. The resonance conditions for the frequency-dependences with field out-of-plane ($f_\perp$) and in-plane ($f_\parallel$) read:

$$f_\perp = \frac{\gamma}{2\pi}\sqrt{\left(H - 4\pi M_{eff} - \frac{4}{3}\frac{K_4}{M}\right)\left(H - 4\pi M_{eff} - \frac{4}{3}\frac{K_4}{M} + 2\frac{K_{2\parallel}}{M}\right)}, \tag{S22}$$

$$f_\parallel = \frac{\gamma}{2\pi}\sqrt{\left(H - \frac{2K_{2\parallel}}{M}\cos 2(\varphi - \varphi_u)\right)} \times \sqrt{\left(H + 4\pi M_{eff} - \frac{K_4}{M} - \frac{2K_{2\parallel}}{M}\cos^2(\varphi - \varphi_u) - 2\left(\frac{K_4}{M}\right)^2 \cos^2(3\varphi)\right)}$$

(S23).

Examples of angle- and frequency-dependent FMR measurements with out- and in-plane configuration of the magnetic bias field are shown in Figure S2.

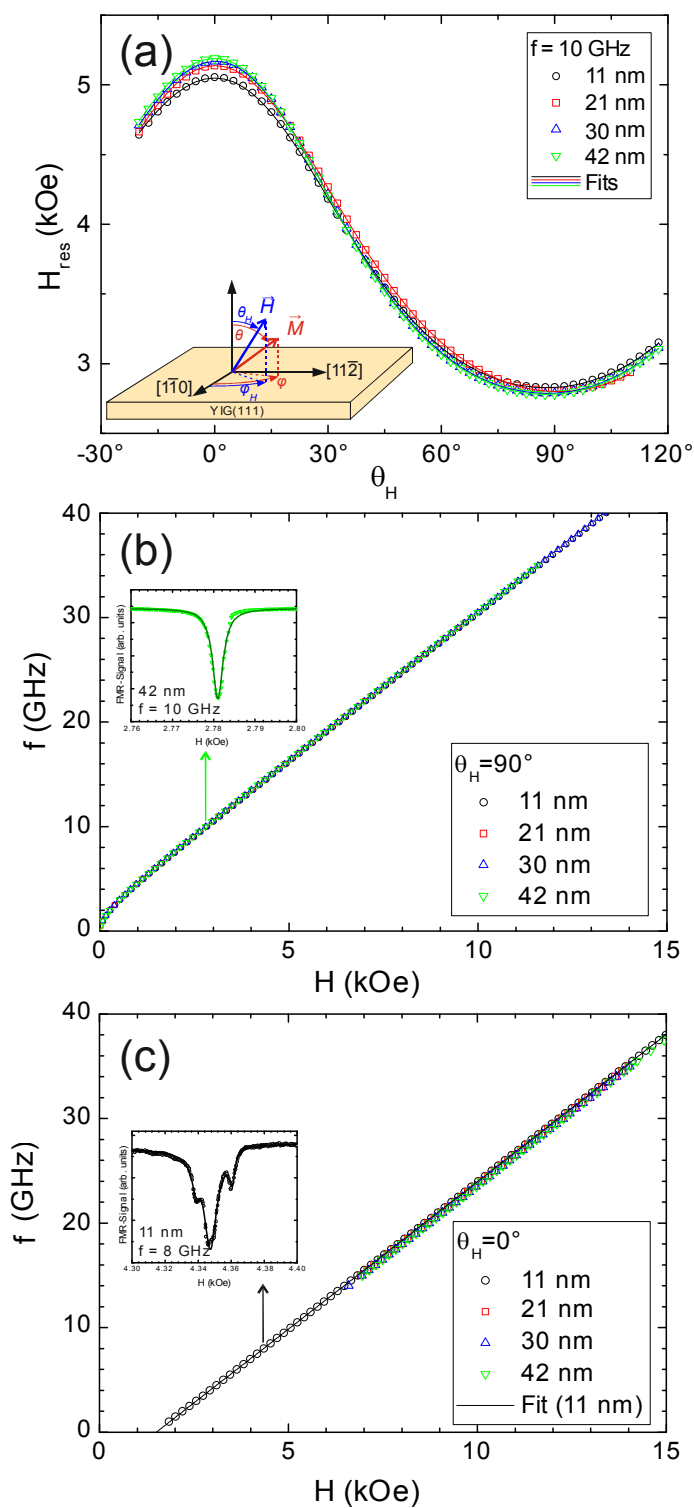

Figure S2. (a) Polar angular dependencies of the FMR measured at $f = 10$ GHz. The inset shows the FMR coordinate system. Solid lines are fits according to the resonance equation. (b) Frequency dependencies of the resonance field measured with field in-plane and (c) out-of-plane. The solid black line is a fit to the 11 nm dataset. Other fit curves have been omitted for visual clarity. Insets show FMR spectra and the indicated positions including Lorentzian fits.